\documentclass[11pt]{article}
\makeatletter
\RequirePackage[T1]{fontenc}
\RequirePackage[utf8]{inputenc}
\RequirePackage[english]{babel}
\RequirePackage{lmodern}
\RequirePackage{microtype}
\RequirePackage{csquotes}
\RequirePackage[a4paper,top=2.35cm,bottom=2.35cm,left=2.55cm,right=2.55cm,footskip=1.15cm]{geometry}
\RequirePackage{amsmath,amssymb,amsthm,mathtools}
\RequirePackage{booktabs,tabularx,longtable,array,multirow}
\newcolumntype{Y}{>{\raggedright\arraybackslash}X}
\newcolumntype{L}[1]{>{\raggedright\arraybackslash}p{#1}}
\RequirePackage{graphicx}
\RequirePackage{enumitem}
\RequirePackage{xcolor}
\definecolor{elpblack}{gray}{0.00}
\definecolor{elpgraydark}{gray}{0.25}
\definecolor{elpgraymid}{gray}{0.55}
\definecolor{elpgraylight}{gray}{0.96}
\RequirePackage[hyperfootnotes=false]{hyperref}
\RequirePackage[nameinlink,noabbrev]{cleveref}
\RequirePackage{titlesec}
\RequirePackage{fancyhdr}
\RequirePackage{setspace}
\RequirePackage{caption}
\RequirePackage{subcaption}
\RequirePackage{url}
\RequirePackage{float}
\RequirePackage{needspace}
\RequirePackage{tikz}
\usetikzlibrary{arrows.meta,positioning,calc,fit,backgrounds,decorations.pathreplacing,patterns}
\RequirePackage[
 backend=biber,style=authoryear,natbib=true,maxcitenames=2,maxbibnames=20,
 uniquename=false,doi=true,url=true,isbn=false,giveninits=true,dashed=false
]{biblatex}
\defbibheading{bibliography}[\refname]{\section*{#1}}
\hypersetup{hidelinks,pdfborder={0 0 0}}
\setlist{nosep,leftmargin=*}
\titleformat{\section}{\Needspace{7\baselineskip}\large\bfseries}{\thesection}{0.65em}{}
\titleformat{\subsection}{\Needspace{5\baselineskip}\normalsize\bfseries}{\thesubsection}{0.65em}{}
\titleformat{\subsubsection}{\normalsize\itshape}{\thesubsubsection}{0.65em}{}
\fancypagestyle{plain}{\fancyhf{}\fancyfoot[C]{\small\thepage}}
\newtheorem{theorem}{Theorem}[section]
\newtheorem{proposition}[theorem]{Proposition}

\theoremstyle{definition}

\newtheorem{criterion}[theorem]{Criterion}

\theoremstyle{remark}

\newcommand{\paperversion}[1]{\gdef\PaperVersion{#1}}
\newcommand{\paperseries}[1]{\gdef\PaperSeries{#1}}
\newcommand{\papersubtitle}[1]{\gdef\PaperSubtitle{#1}}
\newcommand{\paperaffiliation}[1]{\gdef\PaperAffiliation{#1}}
\newcommand{\papercontact}[1]{\gdef\PaperContact{#1}}

\AtBeginDocument{%
  \providecommand{\PaperVersion}{draft}%
  \providecommand{\PaperSeries}{ForesightFlow Research \textperiodcentered{} Event-Linked Perpetuals}%
  \providecommand{\PaperSubtitle}{}%
  \providecommand{\PaperAffiliation}{}%
  \providecommand{\PaperContact}{}%
  \providecommand{\PaperStatus}{}%
}
\newcommand{\authoritynote}[1]{\begin{center}\begin{minipage}{0.92\linewidth}\small\textit{#1}\end{minipage}\end{center}}
\tikzset{
 pbox/.style={draw=black,rounded corners=2pt,align=center,inner sep=5pt,font=\footnotesize,text width=23mm,minimum height=9mm,fill=white},
 pwide/.style={draw=black,rounded corners=2pt,align=center,inner sep=5pt,font=\footnotesize,text width=31mm,minimum height=10mm,fill=white},
 psoft/.style={draw=black,dashed,rounded corners=2pt,align=center,inner sep=5pt,font=\footnotesize,text width=26mm,minimum height=9mm,fill=white},
 parr/.style={-{Latex[length=2.0mm]},line width=0.65pt,draw=black},
 pdash/.style={-{Latex[length=2.0mm]},line width=0.65pt,dashed,draw=black},
 pbrace/.style={decorate,decoration={brace,amplitude=4pt},line width=0.55pt,draw=black},
 ffbox/.style={pbox},ffsmall/.style={pbox,font=\scriptsize,inner sep=4pt},ffsoft/.style={psoft},ffarrow/.style={parr},ffdashed/.style={pdash}
}
\makeatletter
\renewcommand{\maketitle}{%
  \begin{center}
    {\small\scshape \PaperSeries\par}
    \vspace{0.8em}
    {\LARGE\bfseries \@title\par}
    \ifx\PaperSubtitle\@empty\else\vspace{0.35em}{\large \PaperSubtitle\par}\fi
    \vspace{0.75em}
    {\large \@author\par}
    \ifx\PaperAffiliation\@empty\else\vspace{0.18em}{\small \PaperAffiliation\par}\fi
    \ifx\PaperContact\@empty\else\vspace{0.12em}{\small \PaperContact\par}\fi
    \vspace{0.32em}{\small \@date\par}
  \end{center}
  \ifx\PaperStatus\@empty\else\authoritynote{\PaperStatus}\fi
  \vspace{0.25em}
}
\makeatother

\makeatother
\usepackage{xurl}
\hypersetup{pdftitle={On-Demand Combinatorial Event Markets on Kalshi: Instantiation, Concentration, and Effective Market Breadth},pdfauthor={Maksym Nechepurenko},pdfsubject={Event-Linked Perpetuals --- Kalshi Research Track},pdfkeywords={Kalshi, combinatorial markets, multivariate events, market formation, concentration}}
\paperseries{ForesightFlow Research \textperiodcentered{} Event-Linked Perpetuals \textperiodcentered{} Paper 7.2}
\paperversion{r0.7.10}
\papersubtitle{Instantiation, Concentration, and Effective Market Breadth}
\title{On-Demand Combinatorial Event Markets on Kalshi}
\author{Maksym Nechepurenko}
\paperaffiliation{Research Department of Devnull FZCO, Dubai, UAE}
\papercontact{maksym@devnull.ae \quad ORCID 0000-0002-9515-8841}
\date{September 2026 \textbar{} Version r0.7.10}
\begin{document}
\maketitle
\begingroup
\renewcommand{\thefootnote}{}
\footnotetext{\footnotesize\textit{ELP --- Kalshi Research Track.} This paper forms part of the Kalshi-focused research track within the Event-Linked Perpetuals series.}
\endgroup
\begingroup\footnotesize
\begin{abstract}
Kalshi's multivariate-event architecture produces market objects on demand from exact selected legs. Across a registered seven-day interval, 190 independently validated temporal shards yield 7,611,594 unique REST MVE market tickers after excluding 5,777 boundary-overlap observations; the population was created at an average rate of 1.087 million objects per day, with strong hourly burstiness.

The hierarchy is sharply compressed relative to the market-object count: 5,262,526 exact event keys, three collection keys, 83,701 selected-leg primitives, and 66,344,938 selected-leg occurrences. The largest collection accounts for 7,217,085 objects (94.82 percent), with an effective collection count of 1.11; the effective primitive count is approximately 720. REST and WebSocket are distinct observation surfaces: 1,487,330 created-notification tickers intersect the REST population in only 276,177 exact tickers.

All three observed collection keys have current endpoint confirmation without establishing historical collection or rule versions. The exact structural signature yields 7,611,594 unique signatures with zero collision groups. At retrieval, 2,692,787 objects (35.38 percent) have positive cumulative volume or open interest; all observed 24-hour-volume values are zero and trade-count fields are unavailable. This is current-snapshot activity eligibility, not historical trading. The central result is that a very large on-demand market-object population is generated by a concentrated collection layer and reused primitive vocabulary; economic breadth must be measured at several hierarchical levels rather than by ticker count alone. Exact public endpoint evidence is available for 7,357,576 MVE objects, whereas only 126,806 have an exact determination-to-endpoint pair: terminal-state coverage and lifecycle-path coverage are distinct statistical objects.
\end{abstract}
\endgroup
\noindent\textbf{Keywords:} combinatorial markets; multivariate events; market formation; concentration; Kalshi.\\
\textbf{JEL:} G13, G14, G18.

\section{Introduction}
Combinatorial information markets were proposed to express beliefs about joint and conditional outcomes rather than isolated binary events \parencite{hanson2003combinatorial}. Their empirical scale, however, is easy to misread. A venue can expose a very large number of joint-contract tickers while the underlying primitive set, trader attention, and actual activity remain concentrated.

Kalshi's MVE interface sharpens this distinction. The public API documents collections of eligible events and an authenticated endpoint that creates an individual market from selected legs; the endpoint must be called before that market can be traded or looked up \parencite{kalshiMVECollections2026,kalshiCreateMVE2026}. An observed MVE population is therefore not simply an exchange-curated catalogue. It is at least partly the realized output of platform-mediated or user-mediated requests. Creation intensity is consequently an endogenous demand-and-architecture process.

This observation changes the research question. The paper does not ask how many combinations are mathematically possible, because the full feasible set is generally not public. It asks which combinations are actually instantiated, how they map to collections and primitive legs, how concentrated that instantiation is, and what fraction becomes economically active. The bounded current terminal cross-section measures population structure, while the prospective lifecycle stream identifies creation and subsequent attrition.

The nearest empirical market-formation work studies which uncertainties become listable and how settlement rules shape the resulting contract inventory \parencite{adegbenro2026formation}. Paper 7.2 addresses a different margin inside an already supported combinatorial architecture: which eligible leg combinations are actually instantiated on demand, how much primitive information they reuse, and whether instantiated breadth survives an activity adjustment. The object is realized combinatorial supply, not geographic listing selection or market-maker design.

Recent mechanism-design work develops automated pricing for parlay families and tests those mechanisms on Kalshi data \parencite{parlaymarket2026,moshrefi2026apmm}. Paper 7.2 is complementary: it measures the production population to which such mechanisms might apply. Its unit is the instantiated market object and its exact primitive structure, not the design of a market maker.

\section{Institutional creation mechanism and non-intervention}
Let $\mathcal C$ denote MVE collections, $\mathcal P$ primitive leg identities, and $\mathcal M$ instantiated market objects. For market $m$, let
\[
\ell(m)=\bigl(\text{collection key},\ \text{selected-leg tuple}\bigr)
\]
be the raw documented structure. Normalized signatures are derived only from the exact returned fields and retain the raw tuple. Display-title similarity never creates an edge or declares equivalence.

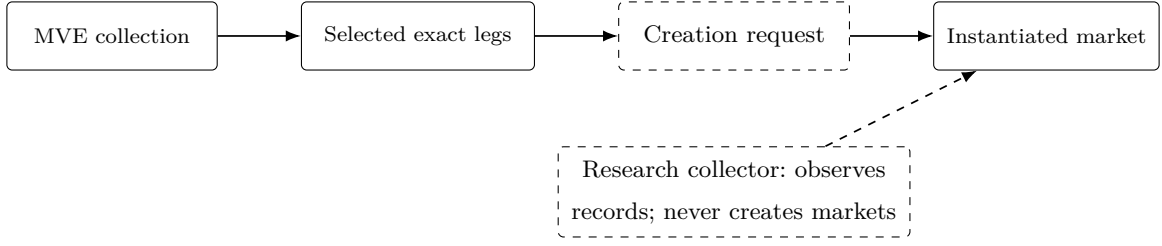
\begin{figure}[H]
\centering
\begin{tikzpicture}[node distance=7mm and 11mm]
\node[ffsmall,text width=25mm] (collection) {MVE collection};
\node[ffsmall,text width=28mm,right=of collection] (legs) {Selected exact legs};
\node[ffsoft,text width=27mm,right=of legs] (request) {Creation request};
\node[ffsmall,text width=27mm,right=of request] (market) {Instantiated market};
\draw[ffarrow] (collection)--(legs);
\draw[ffarrow] (legs)--(request);
\draw[ffarrow] (request)--(market);
\node[ffsoft,text width=43mm,below=10mm of request] (observer) {Research collector:\ observes records;\ never creates markets};
\draw[ffdashed] (observer)--(market);
\end{tikzpicture}
\caption{On-demand instantiation and the non-intervention boundary. The empirical collector observes created markets but does not generate them.}
\label{fig:72-on-demand}
\end{figure}

\begin{criterion}[Non-intervention]
The research data-collection process did not invoke the MVE creation endpoint during the registered acquisition. Any verified intervention would exclude the affected interval from the empirical population.
\end{criterion}

Observed creation is interpreted as realized instantiation under the venue's interface, not as the counterfactual universe of combinations that could have existed. This prevents the research pipeline from contaminating its own outcome and prevents raw creation counts from being mislabeled as exchange listing policy.

\section{Population geometry}
Define the incidence matrix
\[
B_{pm}=1[p\in\ell(m)],
\]
primitive degree $d_p=\sum_m B_{pm}$, and collection multiplicity
\[
M_c=|\{m:m\mapsto c\}|.
\]
For market $m$, define the registered exact production signature
\[
\sigma_{\mathrm{exact}}(m)=
\operatorname{CanonicalIdentity}\!\left[
 e(m),s(m),c(m),\operatorname{sort}_{\mathrm{multi}}\{(e_j,k_j,d_j)\}_{j=1}^{J_m}
\right],
\]
where $e(m)$ is the exact event ticker, $s(m)$ is the market-row series field (absent in the registered population), $c(m)$ is the exact collection ticker, and each selected leg preserves its exact event ticker $e_j$, market ticker $k_j$, and side $d_j$. $\operatorname{CanonicalIdentity}$ is an order-stable representation and $\operatorname{sort}_{\mathrm{multi}}$ sorts lexicographically while preserving duplicate legs. The output MVE ticker, timestamps, activity fields, state, title, and rules are excluded. If $U$ is the number of unique exact signatures, the structural-duplication ratio is
\[
R_{dup}=1-\frac{U}{|\mathcal M|}.
\]
The registered result is an exact production-identity signature, not a claim of economic substitutability. Because the event key is load-bearing, zero collisions under $\sigma_{\mathrm{exact}}$ do not rule out repeated leg multisets across different event identities; a leg-only equivalence class would be a separate robustness object.

Let $w_p=d_p/\sum_q d_q$. Primitive concentration is
\[
H_P=\sum_p w_p^2,
\]
with $1/H_P$ interpreted as the effective number of equally represented primitives, not as independent information or liquidity.

\begin{proposition}[Object count and primitive breadth are not equivalent]
Holding $|\mathcal P|$ fixed, $|\mathcal M|$ can increase without a proportional increase in primitive breadth whenever new condition tuples or repeated combinations reuse the same primitives.
\end{proposition}
The proposition is elementary but consequential: ticker count is not an invariant measure of market breadth.

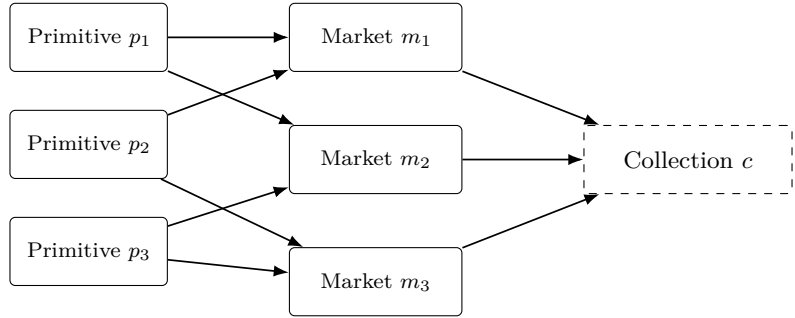
\begin{figure}[H]
\centering
\begin{tikzpicture}[node distance=5mm and 16mm]
\node[ffsmall,text width=18mm] (p1){Primitive $p_1$};
\node[ffsmall,text width=18mm,below=of p1] (p2){Primitive $p_2$};
\node[ffsmall,text width=18mm,below=of p2] (p3){Primitive $p_3$};
\node[ffsmall,text width=20mm,right=of p1] (m1){Market $m_1$};
\node[ffsmall,text width=20mm,below=7mm of m1] (m2){Market $m_2$};
\node[ffsmall,text width=20mm,below=7mm of m2] (m3){Market $m_3$};
\node[ffsoft,text width=24mm,right=of m2] (c){Collection $c$};
\foreach \u/\v in {p1/m1,p2/m1,p1/m2,p3/m2,p2/m3,p3/m3,m1/c,m2/c,m3/c}{\draw[ffarrow](\u)--(\v);}
\end{tikzpicture}
\caption{Primitive--market incidence. Several market objects can reuse the same small primitive set.}
\label{fig:72-incidence}
\end{figure}

\section{Data design}
\subsection{Bounded historical cross-section}
The bounded current terminal cross-section supports terminal population size, collection multiplicity, primitive degree, structural signatures, and terminal-field distributions. It does not identify the historical point process of market creation unless the relevant creation records are independently present. Historical observations are labelled terminal cross-sectional records, not historical lifecycle paths.

\subsection{Prospective inception cohort}
Every MVE creation notification observed during $[T_0,T_0+7d)$ enters the prospective cohort, conditional on documented channel continuity \parencite{kalshiMVELifecycle2026}. Markets are followed through $T_0+14d$. The raw stream also retains older left-truncated markets that emit transitions after $T_0$, but they do not enter full-from-creation denominators.

\subsection{Activity and ordinary benchmark}
Public trades or compact activity summaries are acquired only for the registered MVE cohort and a calendar-matched ordinary inception cohort. The ordinary cohort is a descriptive benchmark. Where feasible, standardized comparisons additionally match underlying event family and creation day; neither design identifies a causal effect of MVE architecture.

\section{Estimands}
Creation intensity is the count $N_M(t,t+\Delta)$ of observed births per registered interval. Concentration is reported through collection and primitive shares, Lorenz curves, Gini/HHI summaries, and top-$k$ shares. No single index is treated as definitive.

The lifecycle funnel is
\[
N_{created}\rightarrow N_{queryable}\rightarrow N_{active}\rightarrow N_{traded}\rightarrow N_{finalized}.
\]
Every stage means ``ever observed'' and carries an observation requirement; reopening and deactivation remain in the full state history.

Let $a_i\ge0$ be a pre-specified activity weight and $\widetilde a_i=a_i/\sum_j a_j$. Define activity-equivalent breadth
\[
B_A=\frac{1}{\sum_i\widetilde a_i^2},\qquad
Q_A=\frac{B_A}{N_{created}}.
\]
$B_A$ is a concentration diagnostic, not a replacement population count. Results are reported for trade count, volume, and capped weights when coverage permits.

For full-from-creation markets, lifespan is $L_i=t_{terminal,i}-t_{created,i}$ with administrative censoring. Creation-to-activation, activation-to-close, and close-to-finalization are reported separately using Paper 7.1 state semantics.

\begin{table}[H]
\centering\footnotesize
\caption{Primary estimands and dependence structure.}
\begingroup\hbadness=10000
\begin{tabularx}{\textwidth}{@{}p{0.22\textwidth}YY@{}}
\toprule
Object & Primary statistic & Inference unit \\
\midrule
Instantiation flow & births per hour/day; burstiness & creation day and collection \\
Collection concentration & top shares, HHI, Lorenz/Gini & collection bootstrap \\
Primitive concentration & degree shares, $H_P$, $1/H_P$ & primitive/collection sensitivity \\
Activity conversion & traded/created and stage funnel & collection \\
Activity-equivalent breadth & $B_A$ and $Q_A$ & collection bootstrap \\
Lifecycle & transition durations with censoring & collection and event \\
\bottomrule
\end{tabularx}
\endgroup
\end{table}

\section{Empirical results from the registered KMVE V1 population}\label{sec:72-results}

\subsection{Population authority and boundary audit}
The primary empirical population is no longer an open-ended archive or a WebSocket message count. It is the exact live REST MVE market-object union whose recorded creation time falls within
\[
[T_0,T_{\mathrm{enroll}})=
[\text{2026-08-15 19:51:29.025 UTC},
 \text{2026-08-22 19:51:29.025 UTC}).
\]
The 190 terminal-validated leaves generate 7,617,371 raw references. Local half-open filtering admits 7,611,594 unique exact tickers and retains 5,777 overlap-envelope observations as excluded boundary evidence. The registered union reports zero exact duplicate groups and zero identity or parent conflicts \parencite{nechepurenko2026kmve}.

\begin{table}[H]
\centering\scriptsize
\renewcommand{\arraystretch}{1.10}
\caption{Registered KMVE V1 population and hierarchy. Market objects, events, collections, and primitives are distinct statistical units.}
\begin{tabularx}{\textwidth}{@{}XrX@{}}
\toprule
Unit & Count & Evidence qualification \\
\midrule
Live REST MVE market objects & 7,611,594 & Exact ticker membership in the seven-day half-open interval \\
Observed exact event keys & 5,262,526 & Exact event-key field in current market rows \\
Observed exact collection keys & 3 & Exact market-row keys; all three independently validated as current identities by read-only GET \\
Current collection-to-series relations & 3 & Parent-derived current observations; not historical market-row fields \\
Selected-leg primitives & 83,701 & Canonical exact primitive identifiers from documented selected-leg fields \\
Selected-leg occurrences & 66,344,938 & Primitive-market incidences; mean 8.72 per market object \\
Current-snapshot activity-eligible & 2,692,787 & Positive volume or open interest at retrieval; not historical trading \\
\bottomrule
\end{tabularx}
\label{tab:72-population-hierarchy}
\end{table}

The REST population corresponds to 1,087,371 market objects per day, or 12.59 per second on average. These rates describe object creation under the registered interface. They do not imply that the same number of distinct source events, traders, or information shocks arrived.

\subsection{Creation intensity and burstiness}
\begin{figure}[H]
\centering
\resizebox{\textwidth}{!}{\begin{tikzpicture}[x=1cm,y=1cm,font=\footnotesize]
\node[anchor=west,text=black] at (0,4.05) {\textbf{Observed MVE market-object creation by UTC hour}};
\draw[black!35] (0,0) rectangle (10.400,3.250);
\draw[black!18] (0,1.083) -- (10.4,1.083);
\draw[black!18] (0,2.167) -- (10.4,2.167);
\draw[black,line width=0.8pt] plot coordinates {(0.0000,0.0800) (0.0619,0.5111) (0.1238,0.5494) (0.1857,0.6690) (0.2476,0.5660) (0.3095,0.4763) (0.3714,0.5381) (0.4333,0.3880) (0.4952,0.2481) (0.5571,0.2182) (0.6190,0.1977) (0.6810,0.1479) (0.7429,0.1083) (0.8048,0.0835) (0.8667,0.0752) (0.9286,0.0748) (0.9905,0.0917) (1.0524,0.1375) (1.1143,0.1969) (1.1762,0.2644) (1.2381,0.3459) (1.3000,0.4664) (1.3619,0.5505) (1.4238,0.5193) (1.4857,0.4941) (1.5476,0.5256) (1.6095,0.3592) (1.6714,0.3877) (1.7333,0.3551) (1.7952,0.2577) (1.8571,0.2364) (1.9190,0.2133) (1.9810,0.1879) (2.0429,0.1822) (2.1048,0.1723) (2.1667,0.1171) (2.2286,0.0902) (2.2905,0.0686) (2.3524,0.0635) (2.4143,0.0771) (2.4762,0.0933) (2.5381,0.1154) (2.6000,0.1458) (2.6619,0.1863) (2.7238,0.2118) (2.7857,0.2085) (2.8476,0.2486) (2.9095,0.2267) (2.9714,0.2140) (3.0333,0.2631) (3.0952,0.3504) (3.1571,0.4021) (3.2190,0.3730) (3.2810,0.3570) (3.3429,0.3557) (3.4048,0.2941) (3.4667,0.3144) (3.5286,0.2722) (3.5905,0.2832) (3.6524,0.1751) (3.7143,0.1245) (3.7762,0.0935) (3.8381,0.0851) (3.9000,0.0905) (3.9619,0.1189) (4.0238,0.1508) (4.0857,0.1812) (4.1476,0.2158) (4.2095,0.2500) (4.2714,0.2745) (4.3333,0.2531) (4.3952,0.2545) (4.4571,0.2526) (4.5190,0.2966) (4.5810,0.4284) (4.6429,0.5305) (4.7048,0.4867) (4.7667,0.5093) (4.8286,0.4175) (4.8905,0.3544) (4.9524,0.3098) (5.0143,0.2577) (5.0762,0.2184) (5.1381,0.1691) (5.2000,0.1299) (5.2619,0.1073) (5.3238,0.0968) (5.3857,0.0988) (5.4476,0.1255) (5.5095,0.1571) (5.5714,0.1935) (5.6333,0.2416) (5.6952,0.3087) (5.7571,0.3761) (5.8190,0.3947) (5.8810,0.3995) (5.9429,0.4124) (6.0048,0.4171) (6.0667,0.4632) (6.1286,0.4801) (6.1905,0.5172) (6.2524,0.4823) (6.3143,0.5585) (6.3762,0.5077) (6.4381,0.3905) (6.5000,0.4029) (6.5619,0.3215) (6.6238,0.2526) (6.6857,0.0977) (6.7476,0.0948) (6.8095,0.1526) (6.8714,0.1384) (6.9333,0.1723) (6.9952,0.2071) (7.0571,0.2504) (7.1190,0.3026) (7.1810,0.3752) (7.2429,0.4445) (7.3048,0.4645) (7.3667,0.4442) (7.4286,0.5073) (7.4905,0.4785) (7.5524,0.5571) (7.6143,0.4986) (7.6762,0.4963) (7.7381,0.4955) (7.8000,0.6465) (7.8619,0.7638) (7.9238,0.9269) (7.9857,0.9155) (8.0476,0.8815) (8.1095,0.7341) (8.1714,0.6063) (8.2333,0.5498) (8.2952,0.4999) (8.3571,0.5839) (8.4190,0.7298) (8.4810,0.9118) (8.5429,1.0360) (8.6048,1.2706) (8.6667,1.4082) (8.7286,1.5911) (8.7905,1.6355) (8.8524,1.9068) (8.9143,1.9308) (8.9762,1.9577) (9.0381,2.1689) (9.1000,2.7795) (9.1619,2.3252) (9.2238,2.4061) (9.2857,2.6445) (9.3476,2.3100) (9.4095,1.8463) (9.4714,1.4195) (9.5333,1.6285) (9.5952,1.3589) (9.6571,1.0003) (9.7190,0.7733) (9.7810,0.7163) (9.8429,0.7588) (9.9048,0.9906) (9.9667,1.4475) (10.0286,2.2175) (10.0905,2.1880) (10.1524,2.8284) (10.2143,3.2111) (10.2762,3.2500) (10.3381,2.9672) (10.4000,2.7590)};
\draw[black!35] (0.0000,0) -- (0.0000,-.09);
\node[anchor=north,rotate=35,text=black!65] at (0.0000,-.13) {Aug 15};
\draw[black!35] (1.4857,0) -- (1.4857,-.09);
\node[anchor=north,rotate=35,text=black!65] at (1.4857,-.13) {Aug 16};
\draw[black!35] (2.9714,0) -- (2.9714,-.09);
\node[anchor=north,rotate=35,text=black!65] at (2.9714,-.13) {Aug 17};
\draw[black!35] (4.4571,0) -- (4.4571,-.09);
\node[anchor=north,rotate=35,text=black!65] at (4.4571,-.13) {Aug 18};
\draw[black!35] (5.9429,0) -- (5.9429,-.09);
\node[anchor=north,rotate=35,text=black!65] at (5.9429,-.13) {Aug 19};
\draw[black!35] (7.4286,0) -- (7.4286,-.09);
\node[anchor=north,rotate=35,text=black!65] at (7.4286,-.13) {Aug 20};
\draw[black!35] (8.9143,0) -- (8.9143,-.09);
\node[anchor=north,rotate=35,text=black!65] at (8.9143,-.13) {Aug 21};
\draw[black!35] (10.4000,0) -- (10.4000,-.09);
\node[anchor=north,rotate=35,text=black!65] at (10.4000,-.13) {Aug 22};
\node[anchor=east,text=black!65] at (-.11,0.0000) {0};
\node[anchor=east,text=black!65] at (-.11,1.0833) {79,134};
\node[anchor=east,text=black!65] at (-.11,2.1667) {158,269};
\node[anchor=east,text=black!65] at (-.11,3.2500) {237,403};
\end{tikzpicture}
}
\caption{Unsmoothed counts of REST-materialized MVE market objects created by UTC hour in the registered seven-day interval. The first and last calendar-hour bins are partial because the scientific interval begins and ends at 19:51:29.025 UTC. The series is a population measurement, not a temporal or causal-effect estimate.}
\label{fig:72-creation-intensity}
\end{figure}
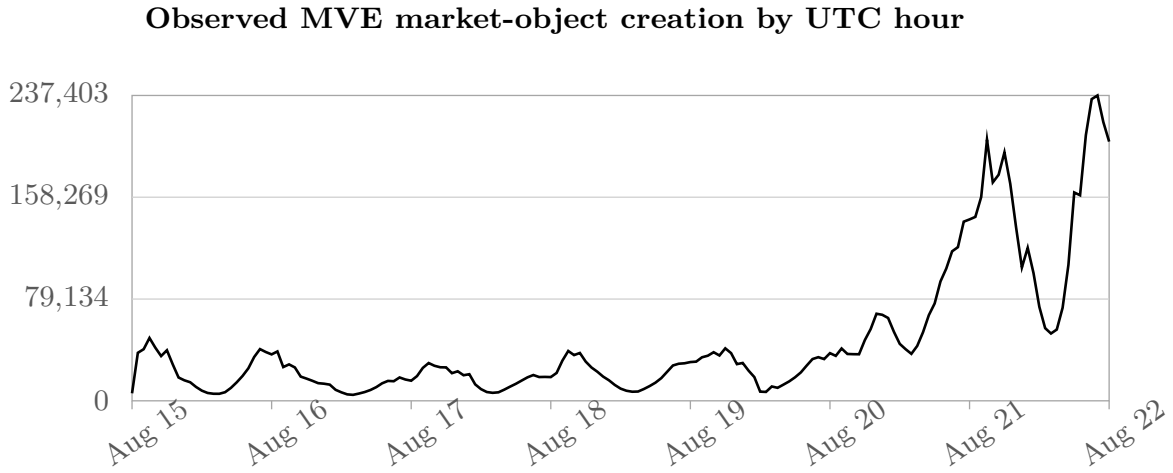

Hourly creation counts vary from 4,640 to 237,403, with a median of 27,404 and a mean of 45,039. The maximum occurs in the hour beginning 2026-08-22 17:00 UTC. The distribution is strongly nonstationary: the final part of the enrollment window contains sustained creation rates above 100,000 objects per hour. Figure~\ref{fig:72-creation-intensity} displays the registered hourly series without smoothing.

This burst structure makes ticker-level independence especially implausible. Primary uncertainty calculations therefore use collection and creation-time clusters, while hourly counts are treated as population measurements rather than draws from a stationary process.

\subsection{Collection concentration}
The exact market-row collection field contains three keys with 7,217,085, 358,400, and 36,109 market objects. Their respective population shares are 94.82\%, 4.71\%, and 0.47\%. The collection HHI is 0.9013, corresponding to an effective collection count of 1.11.

\begin{figure}[H]
\centering
\resizebox{0.9\textwidth}{!}{\begin{tikzpicture}[x=1cm,y=.78cm,font=\footnotesize]
\node[anchor=west,text=black] at (0,3.75) {\textbf{Current collection concentration in the registered seven-day population}};
\node[anchor=east,text=black] at (-.18,2.990) {Collection A};
\fill[black!85] (0,2.800) rectangle (7.1500,3.180);
\node[anchor=west,text=black] at (7.2400,2.990) {7,217,085};
\node[anchor=east,text=black] at (-.18,1.990) {Collection B};
\fill[black!60] (0,1.800) rectangle (5.7906,2.180);
\node[anchor=west,text=black] at (5.8806,1.990) {358,400};
\node[anchor=east,text=black] at (-.18,0.990) {Collection C};
\fill[black!35] (0,0.800) rectangle (4.7514,1.180);
\node[anchor=west,text=black] at (4.8414,0.990) {36,109};
\node[anchor=west,align=left,text width=10.0cm,text=black!65] at (-3.25,-.35) {Bar widths use a log$_{10}$ scale solely to make all three observed collection sizes legible; printed values are exact market-object counts.};
\end{tikzpicture}
}
\caption{Market objects by the three exact collection keys observed in the registered market rows. The log scale prevents the two smaller collections from disappearing. All three keys were independently returned by current read-only collection endpoints; that validation establishes current identity, not historical collection or rule versions.}
\label{fig:72-collection-sizes}
\end{figure}
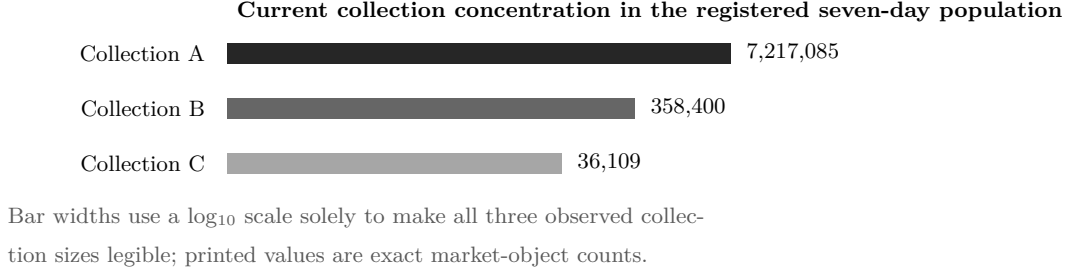

The result is not that Kalshi offers ``three markets.'' It is that 7.6 million market objects in this registered interval are organized under three observed collection identifiers, with one collection accounting for nearly the entire population. This hierarchical concentration is the first load-bearing compression of raw ticker count.

\subsection{Current collection and series identity validation}
Current endpoint validation queried exactly the three observed collection tickers through read-only collection endpoints. All three returned HTTP 200 and exact key parity:
\begin{table}[H]
\centering\footnotesize
\caption{Current collection identity validation. Aliases keep the result table readable. These are current parent observations and are not projected backward into the seven-day historical state.}
\begin{tabularx}{\textwidth}{@{}lXrX@{}}
\toprule
Alias & Current relation & Associated event metadata rows & Evidence status \\
\midrule
C1 & Collection C1 $\rightarrow$ series S1 & 2,008 & GET 200; exact key parity \\
C2 & Collection C2 $\rightarrow$ series S2 & 2,008 & GET 200; exact key parity \\
C3 & Collection C3 $\rightarrow$ series S3 & 2,008 & GET 200; exact key parity \\
\bottomrule
\end{tabularx}
\label{tab:72-current-collection-validation}
\end{table}
\begin{quote}\footnotesize\raggedright
\textit{Exact key map:} C1 \texttt{\nolinkurl{KXMVECROSSCATEGORY-R}} $\rightarrow$ S1 \texttt{\nolinkurl{KXMVECROSSCATEGORY}}; C2 \texttt{\nolinkurl{KXMVECROSSCATEGORY-SHARD1-R}} $\rightarrow$ S2 \texttt{\nolinkurl{KXMVECROSSCATEGORY-SHARD1}}; C3 \texttt{\nolinkurl{KXMVESPORTSMULTIGAMEEXTENDED-R}} $\rightarrow$ S3 \texttt{\nolinkurl{KXMVESPORTSMULTIGAMEEXTENDED}}.
\end{quote}
The market rows themselves contain no non-null series-key field. The three series relations in Table~\ref{tab:72-current-collection-validation} are current parent-derived observations. They validate current parent identity but do not establish which collection or series version governed each market at creation, determination, or settlement. Accordingly, the current-key evidence does not establish historical collection or series identity, even though present key parity is exact. The companion lifecycle interpretation is specified separately in Paper~7.1 \parencite{elp71preprint}.

\subsection{Event multiplicity and primitive reuse}
There are 5,262,526 exact event keys for 7,611,594 market objects, an average of 1.45 market objects per event key. Event size has median one, 90th percentile two, 99th percentile eight, and maximum 4,223. Thus most observed events are associated with one market object, while a small tail supports very large object multiplicity.

Selected-leg fields are present for every included market row. The registered population contains 83,701 exact normalized primitive identifiers and 66,344,938 primitive occurrences. The median primitive appears 19 times, the 90th percentile 780 times, the 99th percentile 14,754 times, and the maximum 1,055,068 times. Primitive HHI is 0.001388, with an inverse-HHI effective primitive count of approximately 720.2. This is the second compression: tens of thousands of primitive identities reduce to a much smaller activity-weighted representation count under degree weights.

The documented identity construction excludes the output MVE ticker, treats selected legs as a sorted multiset, and yields 7,611,594 distinct registered exact signatures; the largest signature group is one. This is a collision result under the exact formula above. It is not evidence that every market represents an independent economic proposition, because exact event identity is itself included and the same primitive vocabulary is heavily reused.

\subsection{REST and WebSocket populations are not the same denominator}
The enrollment WebSocket stream contains 1,487,330 exact tickers with a creation notification and 2,498,640 exact tickers with any lifecycle evidence. Only 276,177 creation-notification tickers and 378,129 all-lifecycle tickers intersect the registered REST creation-time population.

\begin{table}[H]
\centering\footnotesize
\caption{Exact ticker crosswalk between registered MVE observation surfaces.}
\begin{tabularx}{\textwidth}{@{}XrrX@{}}
\toprule
Comparison & Left set & Intersection & Qualification \\
\midrule
WS created notifications vs REST enrollment & 1,487,330 & 276,177 & 18.57\% of WS-created set; 3.63\% of REST set \\
WS all lifecycle vs REST enrollment & 2,498,640 & 378,129 & 15.13\% of WS lifecycle set; 4.97\% of REST set \\
WS created unresolved outside REST & 1,211,153 & --- & No recovery census extrapolates unresolved notifications into REST membership \\
WS left-truncated class & 1,011,310 & 101,952 with REST & No gap extrapolation \\
\bottomrule
\end{tabularx}
\label{tab:72-cross-surface}
\end{table}

The unmatched WebSocket queue is left unresolved rather than converted into fictitious REST parents. Conversely, REST-only tickers are not automatically labelled missed lifecycle messages: the surfaces have different observation rules, and the WebSocket stream contains documented provider gaps. The crosswalk therefore measures surface divergence, not recall of one surface against a presumed gold standard.

\subsection{Current-snapshot activity observability}
Values were parsed under documented numeric validation; null, empty, Boolean, non-finite, and invalid values are non-parseable. All 7,611,594 market rows contain parseable cumulative-volume, 24-hour-volume, and open-interest fields. The four registered trade-count aliases are absent throughout.

\begin{table}[H]
\centering\footnotesize
\caption{Field-specific current-snapshot activity counts. The table describes values observed at retrieval, not trading that necessarily occurred inside the enrollment interval.}
\begin{tabularx}{\textwidth}{@{}XrrrX@{}}
\toprule
Field & Zero & Positive & Unparsable & Interpretation \\
\midrule
Cumulative volume & 4,918,807 & 2,692,787 & 0 & Current cumulative/snapshot field \\
Open interest & 4,940,192 & 2,671,402 & 0 & Current open-interest snapshot \\
24-hour volume & 7,611,594 & 0 & 0 & Zero on every registered row \\
Four trade-count aliases & --- & --- & 0 & Fields unavailable on every row \\
\bottomrule
\end{tabularx}
\label{tab:72-activity-fields}
\end{table}

The registered eligibility expression is
\[
1[\text{current cumulative volume}>0\ \lor\ \text{current open interest}>0\ \lor\ \text{available trade count}>0].
\]
Because no trade-count alias is available, the union contains 2,692,787 market objects, or 35.38\% of the REST population. The union equals the positive-volume set; 2,671,402 members also have positive current open interest. The result is labelled \emph{current-snapshot activity-eligible}. It does not identify the time of activity, the number of trades, or whether activity occurred during the registered seven-day interval, and it is not used as a historical traded-market rate.

\subsection{Downstream lifecycle bridge}
The final 14-day follow-up does not alter the registered KMVE population, its 190-leaf membership proof, or any hierarchy count. It adds a different measurement layer. Exact public endpoint evidence is available for
\[
7{,}237{,}993+119{,}212+371=7{,}357{,}576
\]
MVE market objects: 7,237,993 are REST-only exact endpoints, 119,212 have agreeing WebSocket and REST evidence, and 371 are WebSocket-only exact endpoints. A further 254,018 have no exact endpoint by the registered follow-up boundary.

Exact lifecycle-path evidence is much narrower. Only 126,806 MVE tickers have an exact determination-to-endpoint pair. Thus approximately 96.66 percent of the registered market-object population has an exact endpoint, while the exact paired-clock subset is approximately 1.67 percent. These percentages have different denominators and meanings: the first measures public terminal endpoint availability; the second measures availability of both venue clocks. Neither number is a count of independent events or traders.

This bridge reinforces the paper's hierarchical claim. Market-object population, terminal endpoint coverage, exact two-clock coverage, and full path/revision completeness are separate layers. The companion lifecycle paper owns the duration, censoring, gap, timer, and revision interpretation. Paper 7.2 uses the follow-up only to show that architecture and lifecycle observability cannot be compressed into one ticker-level funnel.

\subsection{What the empirical results identify}
The registered evidence establishes the size and hierarchy of the market-object population, its exact temporal intensity, the concentration of observed collection and primitive identifiers, and the non-equivalence of REST and WebSocket ticker populations. It does not establish user demand, welfare, independent information production, historical trading for every object, or the theoretical feasible combination space.

\section{Registered empirical contrasts}
\textbf{C1 --- Concentration.} Quantify the share of instantiated markets accounted for by the largest collections and primitive legs. The manuscript reports the full distribution rather than a vague heavy-tail label.

\textbf{C2 --- Instantiation-to-activity conversion.} Estimate the probability that a created MVE becomes queryable, active, and traded under complete cohort-linked activity coverage. Missing activity surfaces remain unclassified rather than being treated as untraded.

\textbf{C3 --- Raw versus activity-equivalent breadth.} Report $Q_A$ under several pre-specified weights and caps. Compression is interpreted as concentration of economic use, not proof that inactive objects are valueless.

\textbf{C4 --- Lifecycle standardization.} Compare MVE and ordinary inception paths after standardizing by creation day and broad event family where support exists. The comparison is descriptive and reports common-support attrition.

\textbf{C5 --- Burst and hierarchy sensitivity.} Repeat concentration and creation summaries after removing the largest collection, under raw versus normalized signatures, and across collector-continuity strata.

Uncertainty is obtained by block/bootstrap resampling at collection level, with creation-day blocks for temporal burstiness. Individual MVE tickers within one collection are not treated as independent observations.

\section{Validity and reproducibility}
The results rely on non-intervention, audited MVE lifecycle continuity, exact collection and selected-leg joins, deterministic signature reconstruction, and appropriate denominators for activity claims. Population, boundary, deduplication, event identity, selected-leg, and exact-signature evidence support the stated results. Current collection and series identity, current-snapshot activity observability, historical exact recovery, and rule or source history remain limited to their respective evidentiary scope; none supports a stronger claim than the evidence permits. A possible combinatorial denominator $|\Omega_c|$ is reported only if the venue documentation fully specifies it; otherwise no instantiation ratio is inferred.

The evidence hierarchy distinguishes acquisition integrity, bounded terminal scale, prospective creation flow, collection and primitive geometry, activity observability, lifecycle, and robustness.

\section{Limitations}
A seven-day inception window may capture event-specific bursts. Creation is endogenous to user/platform requests and cannot be interpreted as latent demand from all traders. The public API may not reveal every collection constraint. The three independent collection queries are current observations and do not prove historical collection or rule versions. The exact structural signature includes event identity and therefore does not establish leg-only semantic substitutability. Activity fields are current snapshots without activity timestamps, and historical trading remains unidentified. Results describe instantiated architecture, not welfare, trader identity, or the full theoretical state space.

\section{Conclusion}
The seven-day Kalshi MVE population is enormous at the market-object level: 7,611,594 exact REST tickers. Yet the same evidence immediately shows why ticker count is not economic breadth. The objects map to 5,262,526 observed event keys, three observed collection keys, and 83,701 selected-leg primitives. One collection accounts for 94.82 percent of all market objects, and primitive degree concentration implies an effective primitive count near 720 under the registered weighting. REST creation-time membership and WebSocket notification sets overlap only partially, so the venue's public architecture cannot be represented by one ticker universe.

These results make on-demand instantiation a substantive market-architecture phenomenon rather than a storage anomaly. The main empirical contribution is hierarchical: object creation, event identity, collection structure, primitive reuse, observation-surface membership, exact production signatures, and current activity are different layers. Current validation confirms all three collection identities, identifies their current parent series, and confirms that the output MVE ticker is excluded from the signature. The remaining qualifications are historical rather than numerical: current parent and activity fields are not projected backward, and unresolved WebSocket-only observations are not converted into REST membership. The completed downstream follow-up now confirms that final endpoint evidence can be nearly population-wide while exact two-clock lifecycle paths remain much narrower. That result does not redefine the seven-day population; it validates the paper's insistence that market objects, endpoints, and paths are different measurement layers.

\appendix
\section{Data-field provenance}
The analysis distinguishes collection identity, selected-leg tuples, creation time, lifecycle state, current activity, and terminal state. Display text is descriptive only.

\section*{Data and Code Availability}
The registered population and hierarchy data are available as the \emph{Kalshi Multivariate Event Market Materialization Dataset (KMVE)}, Mendeley Data, V1, doi: \href{https://doi.org/10.17632/fn65786cg6.1}{10.17632/fn65786cg6.1}. The downstream lifecycle materials are summarized here only to distinguish endpoint coverage from exact path coverage. No raw authenticated payloads are included.

\section*{Generative AI Disclosure}
OpenAI ChatGPT and Codex were used for editorial and technical assistance during manuscript preparation. The author made all substantive research decisions, reviewed the final manuscript, and assumes full responsibility for its contents.

\section*{Funding}
This research received no external funding.

\section*{Competing Interests}
The author is affiliated with the Research Department of Devnull FZCO and leads the ForesightFlow research programme. No external sponsor influenced the research design, analysis, interpretation, or decision to publish. The article does not evaluate a commercial product or make investment recommendations.

\printbibliography[heading=bibliography,title={References}]
\end{document}